\documentclass[12pt,letterpaper]{article}
\pdfoutput=1
\usepackage{multirow}

\usepackage{graphicx,array}
\usepackage[dvipsnames,table]{xcolor}
\definecolor{darkblue}{rgb}{0,0.1,0.5}
\definecolor{darkgreen}{rgb}{0,0.5,0.2}
\definecolor{darkred}{RGB}{153,26,0}
\usepackage{ulem}
\definecolor{seablue}{rgb}{0,0.2,0.6}
\usepackage{latexsym}
\usepackage{amssymb, amsmath}
\usepackage{bm}        % for math
\usepackage[numbers,sort&compress]{natbib}
\usepackage{bm,relsize}
\usepackage{slashed}
\definecolor{viola}{RGB}{134,41,198}
\usepackage{mathrsfs}
\usepackage{hyperref} %Automatically links \label and \ref commands; Always load last
\hypersetup{
    colorlinks=true,       % false: boxed links; true: colored links
    linkcolor=darkblue,          % color of internal links
    citecolor=BrickRed,        % color of links to bibliography
    filecolor=magenta,      % color of file links
    urlcolor=MidnightBlue           % color of external links
}
\usepackage[all]{hypcap} %Link navagates to top of figure instead of caption (below fig)
\usepackage{subcaption}

\usepackage[font=small,labelfont=bf,labelsep=period]{caption}

\newcommand{\TeV}{\mathrm{TeV}}
\newcommand{\Mpl}{M_{\rm Pl}}

\newcommand{\be}{\begin{equation}}
\newcommand{\ee}{\end{equation}}

\definecolor{verde}{cmyk}{0.4,0,0.8,0.05}

 \date{\today}

\begin{document}

%%%%%%%%%%%%%%%%%%%%%%%%%%%%%%%%%%%%%%%%%%%%%%%%%%%%%%%%%%%%%%%%%%%%%%%%%%
\begin{flushright}

\end{flushright}
\vspace{.6cm}
\begin{center}
{\LARGE \bf Dark Photon Dark Matter\\ \vspace{3mm} without Stueckelberg Mass}\\
\bigskip\vspace{1cm}
{
\large Michele Redi and Andrea Tesi
}
\\[7mm]
 {\it \small
INFN Sezione di Firenze, Via G. Sansone 1, I-50019 Sesto Fiorentino, Italy\\
Department of Physics and Astronomy, University of Florence, Italy
 }

\vspace{2cm}

\centerline{\bf Abstract} 
\begin{quote}
We study the scenario of dark photon Dark Matter where the mass is generated through the Higgs mechanism rather than the constant Stueckelberg mass. 
In this construction the dark sector contains necessarily extra degrees of freedom and interactions that lead to non-trivial dynamics including thermalization,
phase transitions, cosmic string production. As a consequence the predictions of Stueckelberg theories are vastly modified, strongly depending on the couplings
to curvature and on the scale of inflation $H_I$ compared to the scale $f$ of spontaneous symmetry breaking. We find in particular that only in extreme regions of parameter space 
the phenomenology of Stueckelberg dark photon is reproduced. These scenarios are strongly constrained by isocurvature perturbations unless the dark sector is approximately Weyl invariant.

\end{quote}
\end{center}

\vfill
\noindent\line(1,0){188}
{\scriptsize{ \\ E-mail:\texttt{  \href{mailto:michele.redi@fi.infn.it}{michele.redi@fi.infn.it}, \href{andrea.tesi@fi.infn.it}{andrea.tesi@fi.infn.it}}}}
\newpage
\tableofcontents

\section{Introduction}

One of the most elegant production mechanism of Dark Matter (DM), at work even for free fields, is generation from inflationary dynamics \cite{Ford:1986sy,Chung:1998zb}.
In Ref. \cite{Graham:2015rva} Graham-Mardon-Rajendran (GMR) studied the dynamics during inflation of a massive vector  field, known as dark photon,
obtained through the so called Stueckelberg mechanism, i.e. adding a constant mass term to the action of massless U(1) gauge theory. 
Remarkably this leads to favourable conditions for DM as the vector field can be populated 
from inflationary fluctuations while avoiding dangerous isocurvature fluctuations that strongly constrain these types of scenarios \cite{Chung:2004nh}. 
This mechanism can reproduce the DM cosmological abundance for masses as low as $10^{-6}$ eV of great experimental interest \cite{Fabbrichesi:2020wbt,Caputo:2021eaa}, if the scale of inflation is large.

The Stueckelberg vector model relies on the peculiarity of the spontaneous symmetry breaking of U(1) global symmetries, where the associated Nambu-Goldstone boson is not interacting at leading order in the derivative expansion. Due to this technical fact adding a mass $M$ for a U(1) gauge boson does not spoil renormalizability of the theory. 
This should be compared with the non-abelian case where adding the mass leads to a high energy inconsistency or equivalently to a cutoff determined by the coupling $g$ of order $\Lambda \sim M/ g$ -- the theory requires a completion not far above the mass of the particle. The same conclusion would also apply to the U(1) case if higher derivate terms are included but these are uniquely determined so the strong coupling scale can in principle be pushed to large values.

In Ref. \cite{Reece:2018zvv}  doubts on the  consistency of the model with quantum gravity were put forward.
In string theory realizations no light dark photon has been found compatible with a large cut-off of the effective theory \cite{Goodsell:2009xc}.
The Stueckelberg action describes three degrees freedom and any addition of interactions would spoil the nice features of the theory, see \cite{Ahmed:2020fhc,Kolb:2020fwh,Kribs:2022gri,Clough:2022ygm}.
While none of these arguments is conclusive it seems natural to explore the possibility that the mass of a weakly coupled spin-1 field
is generated through the standard Higgs mechanism. 

In this work we study the consequences of this hypothesis for the production of DM.
We find in general that the predictions differ vastly from the ones of GMR and  are very model dependent.
Only in narrow regions of parameters we can identify the dynamics of Stueckelberg massive vector, as the presence of the Higgs field modifies the dynamics and 
generates new contributions to the abundance. 
Other works, including \cite{Firouzjahi:2020whk,Salehian:2020asa,Nakayama:2021avl} have considered 
dark photon DM obtained through the Higgs mechanism with different production mechanisms. In this work we consider the inflationary production of dark photons originating from the Higgs mechanism neglecting in particular parametric resonance effects that either require very small couplings \cite{Dror:2018pdh} or extra degrees of freedom \cite{Agrawal:2018vin,Co:2018lka,Bastero-Gil:2018uel,Moroi:2020has,Co:2021rhi}.

From a phenomenological point of view a very interesting feature of the GMR scenario is the suppression of isocurvature perturbations on cosmological scales.
This arises non-trivially through the interplay of longitudinal and transverse polarizations of the massive vector field. The addition of the Higgs scalar 
changes this conclusion leading to strong constraints from CMB in large regions of parameters. This problem can be structurally solved
if the scalars have conformal couplings to the curvature. 

The paper is organized as follows. In section \ref{sec:model} we describe the abelian Higgs model that realizes the dark photon 
through the Higgs mechanism. In \ref{sec:breaking} we study the dark photon inflationary 
scenario where the U(1) symmetry is broken during inflation and never restored in the thermal history of the universe.
In section \ref{sec:restoration} we consider the post-inflationary scenario where the U(1) symmetry gets broken after inflation. 
This crucially depends on the coupling to curvature $\xi$ and leads to dark photon dark matter through decays of cosmic strings, as discussed in \cite{Long:2019lwl},
or thermal freeze-out. After the conclusions technical appendixes follow on the inflationary production of scalar and vectors and on
the computation of isocurvature perturbations in these scenarios.

\vspace{0.7cm}
{\bf Note added:} While this paper was in the final stages of preparation Ref. \cite{Sato:2022jya} appeared on the archive with a different
realization of vector dark matter through the Higgs mechanism where the U(1) symmetry is broken during inflation but the dark photon mass
is not constant.

\section{Vector Dark Matter with the Higgs mechanism}
\label{sec:model}

We will consider the simplest realization of spin-1 dark matter obtained through the Higgs mechanism: a dark abelian Higgs model.
This dark sector contains a U(1) gauge field, $A_\mu$, and a complex scalar $\Phi$ with unit charge described by the renormalizable lagrangian
\be\label{eq:model}
\mathscr{L}_D= -\frac 1 4 F_{\mu\nu}^2 + |D_\mu \Phi|^2 -\lambda \left(|\Phi|^2- \frac{f^2}2 \right)^2+ \xi |\Phi|^2 R\,.
\ee
where importantly we have included a coupling to the Ricci scalar.  Such coupling is crucial during inflation but does not affect the low energy dynamics of DM.
Two values of the coupling appear particularly motivated: for $\xi=0$, minimal coupling, the free action of the scalar is invariant under shift of $\Phi$, while 
for $\xi=1/6$ the action becomes Weyl invariant, i.e. invariant under rescaling of the metric $g_{\mu\nu}(x) \to \Omega(x) g_{\mu\nu}(x)$.
While our discussion will be framed in the context of the abelian Higgs model we expect the conformal coupling to capture 
the dynamics of  other scenarios. For example in a dynamical realization of symmetry breaking through strong dynamics
the theory is Weyl invariant so that we expect  features similar to the scalar with conformal coupling, $\xi=1/6$.

At tree level the minimum in flat space the minimum of the potential is at $\Phi=f/\sqrt{2}$, and the spectrum consists of a massive vector boson $A$ and a radial mode  $\phi$ with masses\footnote{When the quadratic term in eq. (\ref{eq:model}) is negligible the U(1) is broken by the Coleman-Weinberg mechanism \cite{Coleman:1973jx}. In such a case the potential has an approximate classical scale invariance and the quartic acquires a logarithmic modulation from running, $\lambda\to \beta_\lambda \log (\phi/(e^{1/4}f))$. The effective quartic coupling is $\beta_\lambda=3g^4/(8\pi^2)$ equal to the contribution to the quartic beta function from the gauge field. The mechanism works when the tree-level parameters are such $\lambda\ll g^4$, so that loop effects overcome tree-level ones. When symmetry is broken radiatively $\lambda \propto g^4$, and the radial mode is the lightest state in the dark sector. This latter scenario is more predictive because both the vector and scalar masses are set by the gauge coupling, $M_\phi^2=3g^2/(8\pi^2)M_A^2$, reducing effectively the model to a 2-parameter space.} 
\be\label{eq:masses}
M_A = g f\,,\quad M_\phi=\sqrt{2 \lambda} f\,.
\ee
These two massive states are both potential DM candidates, if the sector is completely secluded from the SM. 
The massive U(1) gauge boson $A$ is stable thanks to charge conjugation. On the contrary, the radial mode is only stable if the decay $\phi\to AA$ is kinematically forbidden, otherwise it decays to the vectors with a rate $\Gamma\approx (M_\phi/f)^2 \times M_\phi/(8\pi)$. It follows then that for $\lambda < 2 g^2$ the radial mode is stable and it becomes the lightest state for $\lambda< g^2/2$.  In all cases, we only work at weak coupling, that is $M_{\phi,A}\ll f$.

Let us now discuss non-gravitational interactions with the SM.  
At the renormalizable level one can include the Higgs portal coupling and the kinetic mixing,
\begin{equation}
|\Phi|^2 |H|^2\,,~~~~~~~~~F_{\mu\nu}B^{\mu\nu}
\end{equation}
The mixing with hypercharge allows the dark photon to decay so it should be extremely small for the dark photon to be a DM candidate. 
This could be naturally suppressed if the  SM gauge couplings unify into a simple group such as SU(5).

The coupling with the SM Higgs cannot be forbidden through any symmetry of the theory.
This does not modify the discussion of the stability of dark photon but allows the $\phi$ to decay to SM and could lead to thermalization 
or freeze-in production of the dark sector.  Demanding that no thermalization takes place throughout the history of the universe through the Higgs portal, $\lambda_{\phi H}|H|^2 |\Phi|^2$, requires $\lambda_{\phi H} < 10^{-10}$. We will assume in what follows that the dark sector does not thermalize with the SM so that its abundance is dominated
by inflationary dynamics. We will also focus on the abundance of dark photons since $\phi$ that is typically heavy remains cosmologically unstable.

Admittedly the smallness of the Higgs portal coupling is not attractive feature of the dark abelian Higgs model. This could be remedied 
if the U(1) gauge symmetry is broken dynamically. This can be realized with a dark QCD-like sector with fermions
that have chiral charges under the dark photon. In this case the scalar field $\Phi$ is replaced by the fermion  condensate
$\langle \bar{\Psi}\Psi\rangle$. Fermions enjoy Weyl invariance in the massless limit so that results similar to $\xi=1/6$ are
expected in this case. A more detailed study of Weyl invariant sector populated during inflation will be studied elsewhere.

\subsection{Phases of the theory}

\begin{table}[h]
\begin{center}
\begin{tabular}{ r | c | c }
  & $H_I > f$ & $f > H_I$ \\ \hline
$\displaystyle \xi \neq 1/6 $& complex scalar production & scalar and vector production (if $M\ll H_I$)\\ 
$\displaystyle \xi = 1/6 $& Weyl invariance& only vector production (if $M\gg H_I$) \\  
\end{tabular}
\caption{\label{tab:cases}\small Phases of Dark Photon Dark Matter from the Higgs mechanism.}
\end{center}
\end{table}

The abelian Higgs model of eq.~\eqref{eq:model}, despite its simplicity, displays a variety of behaviours 
that differ from the Stueckelberg dark photon of Ref.~\cite{Graham:2015rva}. Although the mass scales  are $M_A$ and $M_\phi$, the dynamics is mostly controlled by $f$ and the Hubble scale of inflation $H_I$ that determines whether the symmetry is broken/unbroken during inflation. This is quite different from the Stueckelberg massive vector case, where the only available scale is the mass of the particle.
The cosmology has some similarities with the QCD axion. Roughly there are two regimes,
\begin{enumerate}
\item $H_I> f$: the U(1) symmetry is restored during inflation so that the dark photon is massless.
During SM reheating or during radiation domination the symmetry breaks spontaneously and the dark photon acquires a mass.
\item $H_I< f$: The field $\Phi$ is in the minimum during inflation and the dark photon is always massive.
\end{enumerate}

It is interesting to draw a comparison with the famous case of the QCD axion, where, if the reheating temperature $T_R$ is larger than $H_I$ the Peccei-Quinn symmetry is also restored. As we will see for the dark photon scenario under consideration, if the system thermalizes, the temperature is much lower than the visible one
so that restoration of the symmetry after inflation is very unlikely. We will not consider this possibility further in this work.

The coupling to curvature, that has no effect today, is important in the early universe. During inflation $R=-12 H_I^2$ so that from eq.~\eqref{eq:model} the mass parameter  becomes $M_{\rm eff}^2=-\lambda f^2 +12\xi H_I^2$. Two values of $\xi$ have special interest, 0 and 1/6. $\xi=0$ corresponds to a minimally coupled scalar 
and is associated to a shift symmetry of the complex scalar in the massless limit. $\xi=1/6$ is the conformal coupling to curvature where the action 
becomes invariant under Weyl transformations that rescale the metric. Weyl invariance is explicitly broken by the mass term $\lambda f^2$  and radiative corrections.
Note that for positive $\xi$ the curvature coupling generates a stabilizing potential around the origin.

The inflationary production strictly depends on $\xi$.
For $\xi=1/6$ approximate Weyl invariance implies that the scale factor can be eliminated from the equations of motions so that no significant inflationary production exists.
For $\xi=0$ the scalar perturbations are copiously produced giving rise to a classical expectation value for the field on
superhorizon scales. This is true both in the broken and unbroken phases.
Note that if the symmetry is broken during inflation the corresponding Nambu-Goldstone boson is always minimally coupled
thanks to the shift symmetry and thus it is strongly produced. In the abelian Higgs model this becomes the longitudinal mode of
massive vector boson that drives its production.

The main possibilities are summarized in table \ref{tab:cases}.

\begin{figure}[t]
\begin{center}
\includegraphics[width=.75\textwidth]{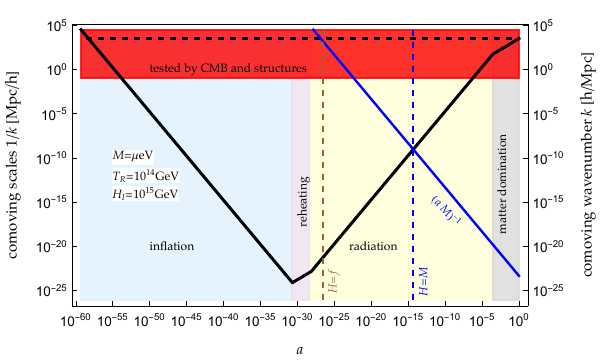}~
\caption{\label{fig:diagram}\small Size of the co-moving Hubble radius $(a H)^{-1}$, solid black line, as compared to the other scales of the model. We also show $H(t)=f$, which is relevant for the case $ H_I > f$.}
\end{center}
\end{figure}

We will make several simplifying assumption throughout. The scale of inflation is taken to be constant during the
visible e-foldings of inflation. Most of the results do not depend on reheating but where they do, at cost of precision, 
we simply treat the reheating as a phase of matter dominance, starting at the end of inflation when the scale factor is $a_e$, with a sudden connection to standard radiation dominance at $a_R$. We also assume that the expansion of the Universe is dominated by the visible sector. The size of the co-moving Hubble radius is shown in figure \ref{fig:diagram}. Depending on the size of $k/a$ compared to Hubble $H$ and to the mass of the particle $M$, we can have different kinematic regimes. 

\subsection{Thermal population}
\label{sec:thermalization}

The interactions in the dark sector are sometime sufficiently strong for the system to thermalize in the relativistic regime.
In such a case, the thermal history of the dark sector is then simply determined by the temperature ratio with the SM,  $r\equiv T_D/T$, defined when the dark sector is relativistic. The actual value of this ratio depends on how much relativistic energy is stored in the dark sector at the onset of thermalization, that is it depends on the production mechanism that populate the dark sector.

The dark sector is unavoidably populated through the SM plasma via gravitational freeze-in, thanks to graviton scattering. Assuming that the energy density thermalizes in the relativistic regime 
one finds \cite{Redi:2020ffc}
\begin{equation}
r_{\rm GR}\equiv \frac {T_D}{T} \sim \left(\frac {T_R}{\Mpl}\right)^{3/4}\,.
\end{equation}
where $T_R$ is the SM reheating temperature.
As we will see, reheating from inflationary fluctuations can also reheat the dark sector.
Gravitational production however can be the dominant source for large reheating temperatures.

\medskip
After thermalization is established, the evolution of the dark sector is controlled by the evolution of $T_D$. When $T_D$ drops below the scale $f$ the U(1) gauge symmetry breaks through a first or second order phase transition and the physical degrees of freedom become the massive dark photon and the dark Higgs field\footnote{For a related discussion in
the context of a non abelian dark sector aka dark QCD see \cite{Garani:2021zrr}.}. Expanding the lagrangian in the fluctuations $\chi$ around the minimum,
\be
\mathscr{L}= -\frac 1 4  F_{\mu\nu}^2 + \frac 1 2 (\partial \chi)^2+ \frac{(f+\chi)^2}{2} (\partial_\mu \theta-g A_\mu)^2 - \frac {\lambda}4 \chi^2(2 f  +\chi)^2
\ee
After the phase transition the energy is mostly transferred to the lightest state. 
Assuming that this is the dark photon this leads to an abundance,
\be
Y_A =\frac{90\zeta(3)}{2\pi^4 g_*} r^3\,
\label{eq:abwdm}
\ee
Here we neglected number changing interaction that could potentially lead to a phase of cannibalism 
in the non-relativistic regime. This would anyway change the abundance only logarithmically.

In the opposite regime the radial mode acts as thermal bath for the dark photons that annihilate until freeze-out.
Their abundance is determined by the Boltzmann equation,
\begin{equation}
\frac {dY}{dT_D}= \frac{\langle \sigma v \rangle s(T)}{H(T) T_D}\big[Y^2- (Y^{\rm eq}(T_D))^2\big]\,,
\end{equation}
where the s-wave annihilation cross-section reads 
\begin{equation}
\langle \sigma v \rangle_{AA\to \phi \phi}= \frac{8\pi \alpha^2}{3M_A^2} \sqrt{1-\frac{M_\phi^2}{M_A^2}}\,.
\end{equation}
The approximate solution of this equation is easily found by noting that this is the standard Boltzmann equation for freeze-out but
with the rescaled cross-section $\langle \sigma v\rangle/r$. Therefore one finds
\be \label{eq:abfreeze-out}
\frac{\Omega_A h^2}{0.12} \approx r \frac{1}{\pi \alpha^2} \left( \frac{M_A}{20\TeV}\right)^2\,.
\ee
Similar arguments can be repeated for the abundance of the Higgs field. In this case however the particle
decays to dark photon if kinematically allowed or possibly to the SM allowing for Higgs couplings.

We wish to emphasize that the thermal contribution is not necessarily the dominant contribution to the energy density in the dark sector. 
For inflationary production non-relativistic modes that re-enter the horizon well after the phase transition might dominate
the energy budget with interesting effects on large scale structures. In particular this leads as we will see to strong isocurvature 
constraints on these models.

\section{$H_I<f$: Higgs phase}
\label{sec:breaking}

We start our discussion considering the case where the U(1) gauge symmetry is spontaneously broken during inflation and never restored
in the thermal history of the universe. 

According to the standard lore the condition ${\rm Max}[H_I,T_{\rm Max}]<f$ imply the symmetry is never restored during or after inflation. However, with couplings to curvature an extra condition exists. Indeed during inflation the effective potential for the radial mode in eq.~\eqref{eq:model} gets a quadratic contribution from the Ricci scalar $R_{\rm inflation}=-12H_I^2$. As such the condition $H_I <f$ can be invalidate for non-zero coupling to the curvature.
Therefore the symmetry is broken provided that
\begin{equation}
\boxed{24 \xi H_I^2 < \lambda f^2=\frac{ M_\phi^2}2\,~~~\&~~~ H_I< f}
\label{eq:xiH}
\end{equation}

Under this hypothesis the field explores the region around the minimum of the potential of eq.~\eqref{eq:model}. 
The physical degrees of freedom are thus a massive vector and a real scalar. 
Working at weak coupling  we always have $M_{A,\phi}\lesssim f$, and the following possibilities are exist.

\subsection{Stueckelberg-like phase: $M_\phi > H_I \gg M_A$}\label{sec:caseA}

In this region of parameter space the abelian Higgs model effectively reproduces the Stueckelberg dark photon of Ref.~\cite{Graham:2015rva} (see also \cite{Ahmed:2020fhc,Kolb:2020fwh}). Since the radial mode mass is larger than $H_I$, its inflationary production is negligible being suppressed exponentially as $\rho_\phi\approx \exp(-2\pi M/H_I)$.  Note that the coupling to curvature $\xi |\phi|^2 R$ is subdominant in this region of parameter space, the field is at the minimum of the potential and never displaced.

Since the only light degree of freedom during inflation is the massive vector the inflationary dynamics is identical to the GMR model. As we summarized in the appendix 
the energy density at late time as function of momentum has the following form
\be
\frac 1 s \frac{d \rho_A^{\rm GMR}}{d \log k}\approx 3\times 10^{-3} \frac{H_I^2}{\Mpl^{3/2}} \sqrt{M_A} \bigg(\frac{100}{g_*(M_\phi)}\bigg)^{1/4} \left\{\begin{array}{lcl} 
\displaystyle   \frac{k_*}{k} & & k\gg k_*\,,\\
%\displaystyle 1  & & k \sim k_*\,,\\
\displaystyle \frac {k^2}{k_*^2} & &  k\ll k_*\,,\\
\end{array}
\right.\,,
\label{eq:PWvector}
\ee
where $k_*\approx a_{\rm eq}\sqrt{H_{\rm eq} M_A}$. Given the shape of the spectrum, the energy density is dominated by the unit logarithmic interval around $k_*$. 
This is shown in figure \ref{fig:PW} where a snapshot of the power spectrum after inflation is shown for massive scalar and vector. 
Integrating over $k$ the dark photon abundance is found
\begin{equation}
\frac{\Omega_A^{\rm GMR} h^2}{0.12}\approx  \sqrt{\frac {M_A}{6 \times 10^{-6}\,{\rm eV}}} \left(\frac {H_I} {10^{14}\,\rm GeV}\right)^2\,.
\label{eq:abGR}
\end{equation}
A very attractive feature of the Stueckelberg dark photon is the fact that the power spectrum is peaked at the intermediate scale $k_*$
and strongly suppressed on cosmological scales thus avoiding  isocurvature constraints from the CMB while predicting 
deviation from cold dark matter at shorter scales \cite{Gorghetto:2022sue,Amin:2022pzv}.

\subsubsection*{Comment on the constraints from weak-gravity conjecture}
The condition $f> H_I \gg M_A\equiv g f$ requires a tiny gauge coupling when the dark photon is realized through the Higgs mechanism. 
Using eq. (\ref{eq:abGR}) one finds
\begin{equation}
\boxed{g <6 \cdot 10^{-29}\left( \frac {10^{14}\, {\rm GeV}}{H_I}\right)^5= 3 \cdot 10^{-11}\left( \frac {M_A}{\rm GeV}\right)^{5/4}}
\end{equation}
For example dark photons of mass $M_A=10^{-5}$ eV  would demand $g< 10^{-28}$ to reproduce the DM abundance through
inflationary fluctuations of the massive vector field.

The necessity of such small couplings raises the question of the consistency of this construction. Indeed in the limit $g\to 0$
the U(1) gauge symmetry would  become global while no exact global symmetries are believed to exist in theories of quantum gravity.

The weak gravity conjecture \cite{ArkaniHamed:2006dz} (see \cite{Harlow:2022gzl} for a recent review) constrains the gauge coupling of U(1) gauge theories. 
In particular the UV cutoff of the theory must go to zero as $g\to 0$ as there are no global symmetries in quantum gravity. 
More precisely the simplest version of the conjecture requires the existence of new states charged under the $U(1)$ symmetry with mass smaller than $\Lambda = g \Mpl$.

In the abelian Higgs scenario, since the U(1) symmetry is broken by the vacuum it is not obvious how the  constraint above and related arguments can be applied.
Indeed the black-holes arguments of  \cite{ArkaniHamed:2006dz} rely on the conserved charges of the theory that however do not exist if the symmetry is spontaneously broken.
One point of view is that the conjecture constrains the sign of the mass term around the origin such that the vacuum breaks the symmetry. 
However given that the symmetry is any case restored at high temperatures it is tempting to apply the bounds that would arise if the cut-off $\Lambda< g f$.

Within this assumption, given that $\Lambda > H$ for consistency of the inflationary theory\footnote{Note that
the constraint is on the Hubble scale rather than the energy density. This can understood as follows: The gravity EFT has the expansion 
$S= \frac {\Lambda^4}{g^2}[1 + \frac {R}{\Lambda^2} + \frac{R^2}{\Lambda^2}+\dots]$. For example in string theory $\Lambda$ should be identified with the string scale and $g$ the close string coupling. This means that the EFT is under control as long as $H< \Lambda$ even though the energy density can be larger than $\Lambda$. This agrees with  intuition from thermal field theory since $H_I/(2\pi)\sim T$.},  imposing that inflation produces the whole abundance of DM (\ref{eq:abGR}) we find,
\begin{equation}
H_I<  10^{10} \,{\rm GeV}\,,~~~~~~~~~~~~~M_A> 50 \,{\rm GeV}
\end{equation}

Contrary to Stueckelberg theories this would imply that no light DM can be generated from inflationary fluctuations. This is not surprising since Higgs and Stueckelberg theories
are not continuously connected. In \cite{Reece:2018zvv} it was argued that in supersymmetric extensions of the Stueckelberg dark photon $\Lambda < \Mpl\times {\rm min}[e^{1/3}\,,\sqrt{M_A/(e \Mpl)}]$ also excluding a light dark photon and a large cut-off.
 
Let us note that dark photons with Stueckelberg masses can be constructed in string theory from abelian forms of the 10d supergravity action.
These realizations are clearly consistent but also do not lead to very light dark photons. For example in \cite{Goodsell:2009xc} it is found that the 
dark photon mass is at least
\begin{equation}
M_A^2 \ge \frac {2\pi}{g_s}\frac {M_s^4}{\Mpl^2}
\end{equation}
where $M_s$ is the string scale to be interpreted as maximum cut-off. 
It would be interesting to study more generally the bounds on dark photon with Stueckelberg masses and from the Higgs mechanism  in string theory. 

\subsection{Light radial mode: $H_I > M_{\phi,A}$}\label{sec:caseB}
Another scenario, that substantially deviates from \cite{Graham:2015rva}, is the one where the radial mode is light during inflation. In order not to restore the symmetry and remain in the Higgs phase, this scenario requires $\xi\approx 0$. This is due to the symmetry breaking condition of eq.~\eqref{eq:xiH}. To emphasize the effect of the coupling to curvature we mention, for example, that the conformal coupling $\xi=1/6$ always restores the symmetry in the regime $H_I > M_\phi$. We thus always consider the $\xi=0$ case in the remainder of this section.

\begin{figure}[t]
\begin{center}
\includegraphics[width=.65\textwidth]{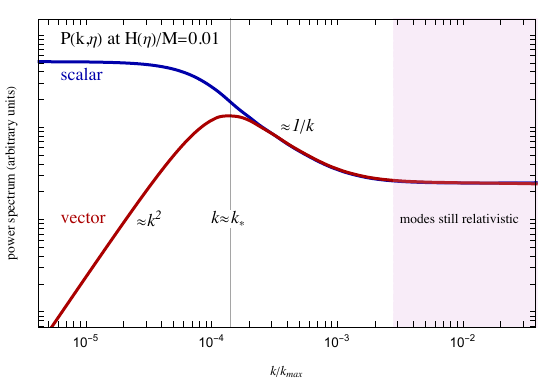}~
\caption{\label{fig:PW}\small Power spectrum of scalar and vector fields produced during inflation obtained following \cite{Kolb:2020fwh,Ema:2018ucl}. The power spectrum is evaluated numerically at a time corresponding to $H=0.01M$ and $k_{\rm max}=a_e H_I$, where $a_e$ is the scale factor at the end of inflation. Here $M$ stands for both the masses of vector and scalar. The typical shape of the non-relativistic modes of scalar and vector field is reproduced, while $k$-mode on the flat part on the right are still relativistic at that time, later on their power will drop as $1/k$. See appendix \ref{app:A} for details. }
\end{center}
\end{figure}

The physical degrees of freedom during inflation are a light minimally coupled scalar and a dark photon. The scalar is produced copiously from inflationary fluctuations (details are given in the appendix). Neglecting decays the power spectrum of the minimally coupled scalar is given by
\be
\frac 1 s \frac{d \rho_\phi}{d \log k}\Big|_{\xi=0}\approx 3\times 10^{-3} \frac{H_I^2}{\Mpl^{3/2}} \sqrt{M_\phi} \bigg(\frac{100}{g_*(M_\phi)}\bigg)^{1/4}  \left\{\begin{array}{lcl} 
\displaystyle  \frac{k_*}{k} & & k\gg k_*\,,\\
\displaystyle 1  & & k \ll k_*\,,\\
\end{array}
\right.\,,
\ee
while the one of the vector is still given by eq.~(\ref{eq:PWvector}). They are displayed together in figure \ref{fig:PW}.

If the scalar is heavier than the vector it rapidly decays.  This contributes to the abundance of dark photons,
\begin{equation}
\frac {\rho_A}s\Big|_{\rm decays} \approx  5\times 10^{-3} \frac{H_I^2}{\Mpl^{3/2}}  \frac {M_A}{\sqrt{M_\phi}}\,\bigg(\frac{100}{g_*(M_\phi)}\bigg)^{1/4}\,\log \frac {k_*}{H_0}\
\label{eq:abdecay3}
\end{equation}
that is smaller than the abundance from inflationary production of the dark photon.

For $M_\phi< 2 M_A$ the scalar cannot decay to dark photons. In this case it could either 
be exactly stable and contribute to the DM abundance or decay to SM, for example through the Higgs coupling.

\subsubsection{Thermalization with a light radial mode}
If the radial mode is light during inflation, interactions between the dark photon and the scalar can be significant and they may lead to thermalization in the dark sector. We then wish to understand how much of the energy produced during inflation can be transformed into thermal energy,
see \cite{Redi:2020ffc} for a related discussion in the context of dark gauge theories.\footnote{
Thermalization of Stuckelberg dark photon with dark electrons was studied in detail in
~\cite{Arvanitaki:2021qlj}. Differently from this this Ref. we consider no light charged states, such that we do not expect the presence of a rich physics in the form of a dark-QED plasma on sub-horizon scales.}

As shown in figure \ref{fig:PW} scalar and vector modes with $k>k_*$ behaves identically. On such scales, modes re-enter the horizon 
while relativistic with momentum $k/a\sim H$. For significant interactions this may lead to thermalization
in the dark sector.  Modes with $k\gg k_*$, at horizon re-entry have a density $d n_k/d\log k \sim H_I^2 k/a_k$, 
for both scalar and vectors given that for such wave-number they behaves in the same way (see appendix). 
On distances  $\sim a/k$, thermalization is determined by scattering rates of order
\be
\frac{d\Gamma}{d\log k} \approx \left(\frac{H_I^2}{(2\pi)^2}\frac{k}{a_k}\frac{a_k^3}{a^3}\right)\times \frac{\alpha_{\rm eff}^3}{(k/a)^2}\,,
\ee 
where we have parametrized the cross-section relevant for thermalization in the relativistic regime as $\sigma_{\rm eff} = \alpha_{\rm eff}^3/p^2$.
We can see that the interaction rate scales as $1/a$, while Hubble redshift as $1/a^2$ in radiation, leading to thermalization if the coupling is not exceedingly small. Moreover, since the energy of the modes are $\approx H$ at re-entry, the above expression at re-entry is $d\Gamma/d\log k |_{\rm re-entry}\approx \alpha^3 H_I^2/H$, much larger than Hubble. 

For modes that re-enter the horizon non-relativistically, $k/a< M$, the situation is different for scalar and vector. As shown in Fig. \ref{fig:PW}
the latter is strongly suppressed in the IR. For the scalar the numerical density does not change by the cross-section goes to constant.
For these reasons thermalization becomes inefficient and modes with small momenta do not thermalize.
For simplicity we will thus assume that the energy density of the thermal bath is determined by the modes that re-enter relativistically.

The computation goes as follows. An energy density of order $H_I^2 H^2/(2\pi)^2$ per Hubble time gets converted into dark thermal energy. 
Under the assumption of instantaneous thermalization at horizon re-entry, the energy density of the dark sector is
determined by the Boltzmann equation \cite{Redi:2021ipn}
\begin{equation}
\frac {d\rho} {dt}+ 4 H \rho=\gamma_1\,,~~~~~~~~~\gamma_1\approx \left(\frac{H_I}{2\pi}\right)^2 H^3 \,.
\end{equation}
We can easily integrate this equation in radiation domination to find
\begin{equation}
\rho(T)\approx T^4 \left(\frac {H_I}{\Mpl}\right)^2 \frac {g_*}{360} \log \frac{\sqrt{\Mpl H_I}}{T}
\end{equation} 
From this we extract the dark sector temperature induced by inflationary production,
\be
r_{\rm inf} \equiv \frac{T_D}{T} \sim  0.3 \left(\frac{g_*}{g_D}\right)^{1/4} \left(\frac{H_I}{\Mpl}\right)^{1/2}\,
\label{eq:TD}
\ee
leading automatically to a dark sector much colder than the SM. 

Notice that even in the extreme case of instantaneous reheating, where the SM temperature reads $T_R\sim \sqrt{H_I M_p}/g_*^{1/4}$, the dark sector is never reheated above the Hubble scale during inflation, $T_D^{\rm max}< H_I$. Therefore, starting in the Higgs phase during inflation, the U(1) symmetry is not restored afterwards. 

Interestingly, however, the dark sector temperature can be larger than the physical masses of the scalar and dark photon, which can behave for a while as a thermal plasma. In such a case the discussion flows along the lines of section \ref{sec:thermalization}. As in that section we have two options

If the dark photon is the lightest state this leads to the abundance in eq.~(\ref{eq:abwdm}).
This is suppressed by a factor $\sqrt{M_A/H_I}$ compared to the abundance from inflationary fluctuations in eq. (\ref{eq:PWvector}). In the opposite regime $M_A> M_\phi$ the thermal abundance of dark photons is determined by thermal freeze-out as in eq. (\ref{eq:abfreeze-out}).
Comparing with eq.~\eqref{eq:abGR} we find that $\Omega_A^{\rm GMR}/\Omega_A^{f.o.}\approx  10^{-14}g^2 H_I^2/f^2 \sqrt{g\, {\rm GeV} /f}$ so that the freeze-out contribution 
can easily dominate in the region of interest $f \gg H_I$. We notice that thermalization occurs only if the couplings are not extremely tiny.

\begin{figure}
\begin{center}
\includegraphics[width=.65\textwidth]{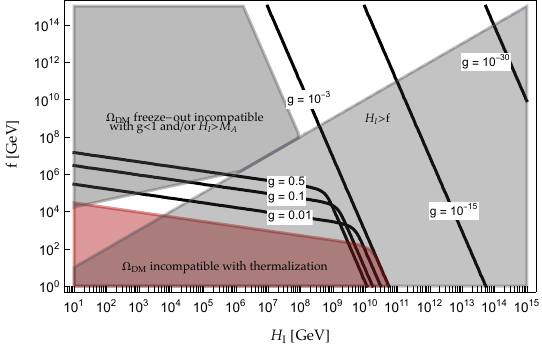}
\caption{\label{symmetryrestorationPT}\small Higgs phase of Dark Photon DM. Black isolines correspond to region where the DM abundance is reproduced. For sizeable gauge coupling the dark sector thermalizes and dark photon can undergo thermal freeze-out, while for tiny coupling only the GMR mechanism is at work.}
\end{center}
\end{figure}

\subsubsection{Isocurvature constraints}
\label{sec:isoHiggs}
The fluctuations of dark photon and radial mode are orthogonal to the ones of the inflaton during inflation.
This produces non-adiabatic perturbations in the matter power spectrum that are strongly constrained by CMB
and large scale surveys. In the Higgs realization of the dark photon strong constraints from isocurvature are back to life. 
For $\xi\approx 0$ (which is necessary to break the symmetry with $f> H_I$) the spectrum of fluctuations of the radial mode
is IR dominated producing a DM population with isocurvature perturbations, and the decay of $\phi$  transfers the unhealthy isocurvature population to dark photons.

An assessment of the bounds from isocurvatures is therefore necessary in dark photon models with a radial mode. We here consider the usual parametrization of isocurvature perturbations defined to the entropy density of photons as
\be
\frac{\delta \rho}{\rho}=\frac{3}{4}\delta_\gamma + \delta_{\rm iso}\,.
\ee
$\delta_{\rm iso}$ is the initial data of stochastic origin, and in our case, given that it is produced by the same mechanism that produces DM it is expected to be $O(1)$. In particular the power spectrum associated to $\delta_{\rm iso}$, $P_{\rm iso}=(2\pi^2)/k^3\times\Delta_{\rm iso}(k)$, is the same power spectrum of DM in our case.
The CMB bound can be cast in terms of the power spectrum of the isocurvature population compared to the total abundance as done in \cite{Planck:2018jri}. We introduce the parameter $\beta_{\rm iso}(k)$ defined as
\be
\beta_{\rm iso}(k)=\frac{\Delta_{\rm iso}(k)}{\Delta_{\zeta}(k)+\Delta_{\rm iso}(k)}
\ee
PLANCK sets limits on $\beta_{\rm iso}(k_c)$ at different $k_c=0.002,0.05,0.1\, \mathrm{Mpc}^{-1}$ ($\Delta_{\rm \zeta}(k_*)=2.1 \cdot 10^{-9}$). We make use of the results for the uncorrelated CDI (axion-I, section 9.4.1 of Ref.~\cite{Planck:2018jri}), where the bounds are computed at $k_c=0.05\, \mathrm{Mpc}^{-1}$, $\beta_{\rm iso}(0.05\, \mathrm{Mpc}^{-1})< 0.038$ at 95\%CL.

\bigskip
The minimally coupled scalar has an IR  dominated power spectrum that in isolation would imply $\beta_{\rm iso}(k_c)\sim 1$ grossly excluded experimentally. However, thanks to decay, the contribution to the abundance of dark photon of eq.~\eqref{eq:abdecay3}  is suppressed by  $\sqrt{M_A/M_\phi}$ weakening the bounds. Avoiding constraints would however demand $M_A< 10^{-10} M_\phi$. 

A different possibility to suppress isocurvature perturbations is realized if the mass of the particle is close to the Hubble scale during inflation \cite{Chung:2004nh,Tenkanen:2019aij}.
We discuss this possibility in detail in the appendix \ref{sec:isocurvature}. The power spectrum of a massive scalar field in de-Sitter reads,
\begin{equation}
P(k)\sim \frac {H_I^2} {2k^3} \left( \frac k {a H_I}\right)^{\frac{2 M^2}{3H_I^2}}\,,
\end{equation}
This implies that for $M\sim H_I$ modes that correspond to cosmological scales probed by the CMB and large scale structure are suppressed.
As we show in the appendix \ref{sec:isocurvature} (see also \cite{Markkanen:2019kpv,Tenkanen:2019aij}), assuming for simplicity that $H_I$ is constant during inflation,
one finds,
\begin{equation}
\boxed{\Delta_{\rm iso}(k) \approx \frac {8M^2}{3H_I^2} e^{- \frac{4 N(k) M^2}{3H_I^2}}}\,.
\label{eq:iso-star}
\end{equation}
where $N(k)$ is the number of e-foldings to the end of inflation when $k/a = H_I$. 
Assuming $N(k_*)\sim 50$ this is compatible with isocurvature bounds if $M_\phi> 0.5 H_I$.

For $ M_\phi< 2 M_A$ a similar constraint applies unless  $\phi$ decays to SM.
In that case isocurvature perturbations can be washed out by thermalization with the SM thermal bath.

\section{$H_I>f$: Coulomb phase}
\label{sec:restoration}

In this regime the U(1) symmetry is restored in the early universe and the dark photon is massless during inflation.
This implies that during inflation the relevant degrees of freedom are the real and imaginary parts of $\Phi$ with same coupling $\xi$ to curvature. 
For perturbative couplings we have $M_\phi< H_I$, so that the complex field $\Phi$ can fluctuate during inflation. By constrast the two helicity states of the dark photon are not produced by inflationary fluctuation thanks to their classical Weyl invariance.

There are however several contributions to the abundance of dark photons in this phase:
\begin{itemize}
\item
After inflation the system undergoes a phase transition where the field relaxes to the symmetry breaking vacuum. 
Through the Kibble mechanism cosmic strings are generated that emit dark photons in the scaling regime of the string network producing an
abundance of dark photons \cite{Long:2019lwl}.

\item
For $\xi= 0$ the minimally coupled complex scalar $\Phi$ is produced by inflationary fluctuations with an energy density of order $\approx H_I^4/(2\pi)^2$.
The modes that re-enter relativistically reheat the dark sector at the temperature (\ref{eq:TD}) leading to a thermal abundance of dark photons.
Modes that re-enter at late time can carry significant amount of energy giving rise to isocurvature perturbations.

\item
At the phase transition an energy density of order $\lambda f^4$ is released that also populates the dark sector.
\end{itemize}

Let us now discuss in detail the various contributions. 

\subsection{Conformal coupling $\xi=1/6$}
\label{sec:restoration-conformal}

During inflation the conformal coupling to curvature acts as a positive mass term. 
The equation of motion for the zero mode of the radial field reads,
\begin{equation}
\ddot \phi+ 3 H_I \dot \phi + 2 H_I^2 \phi + \lambda \phi (\phi^2-f^2)=0
\end{equation}
Since $H_I> f$  we can neglect the instability associated to the potential.
For an initial displacement $\lambda \phi_0^2< H_I^2$ the approximate solution is
\begin{equation}
\phi\sim \phi_+ e^{-2 H_I t }+ \phi_- e^{- H_I t }
\end{equation}
This means that, after few e-foldings of inflation, the field relaxes to the origin effectively restoring the symmetry.
The same conclusion also applies if $\lambda \phi_0^2>H_I^2$. 

Let us note that this mechanism is quite different from symmetry restoration due to inflationary fluctuations 
of size $\delta \phi\sim H/(2\pi)$. For $\xi=1/6$ the system is approximately Weyl invariant
and no inflationary production takes place. This can be seen explicitly by writing the equation of motion for the rescaled field,
$v=a\phi$. Around $\phi=0$ the equation of motion reads
\begin{equation}
v''  + k^2 v-  \lambda \frac{f^2}{H_I^2 \eta^2}  v=0  
\end{equation}
where $\eta$ is the conformal time so that during inflation $a=1/(H_I \eta)$. For a minimally coupled scalar instead the instability is $-2/\eta^2$ that leads to strong production
of modes as they exit the horizon. In this case the instability is marginal and  $\Phi$ is not significantly produced during inflation (see for example \cite{Ema:2018ucl}),
\be\label{eq:restored-conformal}
\frac{d\rho_{\Phi}}{d\log k}\bigg|_{\rm inflation}\approx 0\,.
\ee

The dark sector is thus populated only through the phase transition. 
This has two effects: the release of the energy of the false vacuum and the formation of cosmic strings.

\subsubsection{Dark photons from phase transition}

Assuming that the inflaton only reheats the SM, the dark sector at the end of inflation is empty and in the false vacuum at $\Phi=0$.
During reheating or radiation domination, a phase transition takes place with a release of energy $\Delta V = \lambda/4 f^4$. 
Since in radiation domination $R_{\rm rad}\approx 0$ in this phase the coupling to curvature does not stabilize the potential and the system becomes immediately unstable.
The energy density is transferred immediately to the coherent oscillations of the radial mode that behave as non-relativistic particles.
The abundance is given by,
\be
\frac{\rho_\phi}{s}\bigg|_{\rm P.T.}=\frac{\Delta V}{s(T_R)} =\frac {45}{8\pi^2 g_*(T_R)} \frac {\lambda f^4}{T_R^3}\approx  0.03\frac {\lambda f^4}{(H_R \Mpl)^{3/2}}\bigg(\frac{100}{g_*}\bigg)^{1/4}
\label{eq:abPT}
\ee
The scalar quickly decays to dark photons if kinematically allowed.
Therefore the abundance of dark photon can be obtained from eq. (\ref{eq:abPT}) multiplying by $M_A/M_\phi$.

The above estimate needs to be reconsidered if the phase transition happens during reheating. Since we have considered reheating as a phase of matter domination, the coupling to curvature continues to stabilize the potential after inflation. Using $R_{\rm mat}=-3 H^2$ the instability arises for
\be\label{eq:critical}
H_c=2 \lambda f\,.
\ee
If the phase transition takes place during reheating one finds that the abundance (\ref{eq:abPT}) is reduced 
by the factor $H_R^2/H_c^2$, where $H_R$ is the Hubble scale at the onset of radiation domination.

\subsubsection{Dark photons from the string network}
The abelian Higgs model has a vacuum manifold $S_1\simeq U(1)$ that support topological string solutions. 
When the U(1) symmetry breaks spontaneously through the Kibble mechanism the field rolls in a random way to the minimum 
of the potential producing a network of strings. In the evolution of the system the network quickly approaches a scaling
regime where roughly one string per Hubble volume exists. To maintain the scaling regime energy must be released that in this case
is dominated by the emission of massive dark photons. Dark photon emission from the string network was discussed in Ref.~\cite{Long:2019lwl} and we briefly review the main results here.

In dark abelian Higgs model, the spontaneous breaking of the gauge symmetry gives rise to cosmologically stable strings. Since the phase transitions happens when $H(t)\approx f$, the correlation length will be of the order of $1/H$, corresponding to approximately one string per Hubble volume at the phase transition.  Once the string network attains the scaling solution, the energy density at all time is given by
\be
\rho_{\rm string}(t)=\kappa \frac{\mu_{\rm string}}{t^2}\approx \pi f^2 H(t)^2\,.
\ee
where  $\kappa$ represents the number of string per Hubble volume \cite{villadoro1}, while $\mu_{\rm string}\approx \pi f^2$ is the effective string tension. The scaling is reached when the energy density of long strings is efficiently transferred to closed string loops that oscillates and eventually decay into radiation.  Taking into account the expansion of the Universe, the continuity equation gives
\be\label{eq:radiated-power}
\dot\rho_{\rm string} + 2 H \rho_{\rm string}\approx - \xi\frac{\mu_{\rm string}}{t^3} \,,
\ee
which is the amount of power radiated by the network to sustain the scaling (we have neglected scale violations in $\xi$ and $\mu_{\rm string}$, we refer the reader to \cite{villadoro1,villadoro2} for a detailed discussion). The situation is similar to the QCD axion case, with the difference that our U(1) is gauged, therefore the light quanta are the radial mode and the massive vector. Emission of these particles will allow the network to track the above scaling, and when production of vectors will become inefficient, emission into gravitational waves will be enough to mantain the scaling regime . 

We estimate that production of massive vectors will stop around $H(t)\approx M_A$, and gravitational wave emission will be maximal afterwards. Integrating the power of \eqref{eq:radiated-power}, and taking into account that the vectors are produced with physical momenta $p\approx H(t)$ as long as $H\approx M_A$, we can compute the energy density of vectors produced by strings \cite{Long:2019lwl}. The spectrum at production is flat, but the energy density in vectors today is dominated by the mode produced with $p\approx H\approx M$. In order to make comparison with the results of the previous section, we write the energy over entropy ratio, which is conserved. For dark photon produced via strings is
\be\label{eq:yield-strings}
\frac {\rho_A}s\bigg|_{\rm string} \approx \kappa\, \sqrt{M_A} \, \frac{f^2}{\Mpl^{3/2}} \bigg(\frac{100}{g_*}\bigg)^{1/4}\,.
\ee
From this expression, we see that the contribution to dark photon abundance from the string network is always parametrically larger than the one from the phase transition of \eqref{eq:abPT}. 

Present bounds on cosmic strings are mostly due to the non-observation of gravitational waves from pulsar-timing arrays. Experimental limits are given on the combination $G\mu\lesssim 10^{-11}$ \cite{Long:2019lwl}, where $G$ is the Newton constant.

\subsection{Minimal coupling $\xi=0$}\label{sec:restoration-minimal}

For $\xi=0$ the complex field $\Phi$ experiences inflationary fluctuations.
Since the size of the fluctuation is of order $\approx H_I/(2\pi)$, the field explores the region of the potential dominated by the quartic term.
As we will see this complicates the analysis of inflationary perturbations compared to the free theory but leads to similar qualitative results.
In particular isocurvature perturbations are generated that strongly constrain the scenario.

\subsubsection{Misalignment and inflationary production}

Evolution of the classical field (misalignment) and inflationary field fluctuations can both contribute to the energy density of the scalar.

Let us start discussing the misalignment mechanism in this context \cite{Turner:1983he}. For the sake of simplicity, let us consider the radial direction of $\Phi$ described by a real scalar field with quartic potential
\begin{equation}
\mathscr{L}= \frac 1 2 (\partial \phi)^2  - \frac{\lambda} 4 (\phi^2-f^2)^2\,.
\end{equation}
If the  field value is such that $ \phi > f$, then the quartic term dominates the potential.
The effective mass is
\begin{equation}
M_{\rm eff}^2 = \lambda(3 \phi^2-f^2)\,.
\end{equation}
Assuming $H_I> M_{\rm eff}$ the field remains constant during inflation at the initial value $\phi_0$ and starts to oscillate when $H(a_c)\approx M_{\rm eff}$.
Unlike a massive scalar for a quartic potential the coherent oscillations of the field redshift as radiation in light of the conformal invariance 
of the action so that $\phi=\phi_0 a_c/a$. This behaviour continues until the scale factor reaches $a_*$, where the quadratic term in the potential dominates.
This happens at $\phi\sim f$ so that $a_{c}/a_*= f/\phi_0$. Assuming that the oscillations take place during radiation domination we can compute the corresponding Hubble scale as \begin{equation}
H_*= H_c \frac{a_c^2}{a_*^2}\approx \sqrt{3 \lambda} \frac {f^2}{\phi_0}
\end{equation}
With the aid of eq. (\ref{eq:sH}) we can thus compute the ratio energy to entropy, 
\begin{equation}
\frac {\rho} s\Big|_{\rm mis} \approx \frac{\lambda}{4}\frac{f^4}{s_*}\Big|_{\rm quartic} \approx 0.05  \lambda^{1/4} |f| \bigg(\frac{100}{g_*(M_\phi)}\bigg)^{1/4}\left(\frac {\phi_0}{\Mpl}\right)^{3/2} 
\label{eq:misphi4}
\end{equation}
Let us note that this formula is valid for positive and negative $f^2$. 

Inflationary fluctuations in the field $\phi$ are computed in a similar fashion. The linearized equation of motion reads
\begin{equation}
\ddot \chi + 3 H \dot \chi+ k^2 \chi + \lambda(3  \langle \phi^2\rangle-f^2) \chi=0
\end{equation}
For $M_{\rm eff}\ll H_I$ the potential during inflation is negligible so that each mode at the end of inflation has an amplitude $H_I/(2\pi)$. 
The mode starts to oscillate when $H\sim M_{\rm eff}$ that in turn depends on $\langle \phi \rangle$.
Given that the result (\ref{eq:misphi4}) is linear in the mass at late time we get 
\be
\frac 1 s \frac{d \rho_\phi}{d \log k}\Big|_{\rm inf}\approx 0.005 \lambda^{1/4}|f|  \frac{H_I^{3/2}}{( \Mpl)^{3/2}}   \bigg(\frac{100}{g_*(M_\phi)}\bigg)^{1/4} \left\{\begin{array}{lcl} 
\displaystyle  \frac{k_*}{k} & & k\gg k_*\,,\\
\displaystyle 1  & & k \ll k_*\,,\\
\end{array}
\right.\,,
\label{eq:abquartic}
\ee

\subsubsection{Dark photons abundance}
If the symmetry is unbroken during inflation dark photons are not directly produced by inflationary fluctuation due to Weyl symmetry. 
They can be produced by other mechanisms: from the decays of the field $\phi$, through thermalization and by the network of cosmic strings after when the symmetry is broken.

The contribution from decays inherits the numerical abundance of $\phi$. Assuming that the dark photons are non-relativistic today the contributions of the perturbations (\ref{eq:abquartic}) 
to the dark photon abundance is
\be
\frac {\rho_A} s\Big|_{\rm inf,decays} \approx 0.005 \frac {M_A}{\lambda^{1/4}}  \frac{H_I^{3/2}}{( \Mpl)^{3/2}}   \bigg(\frac{100}{g_*(M_\phi)}\bigg)^{1/4} \log \frac{k_*}{H_0}\,.
\label{eq:abxi1H>f}
\ee

In addition to this population, and similarly to section \ref{sec:caseB}, $\phi$-modes that re-enter the horizon relativistically could thermalize via inelastic processes if the couplings are sizable.
This gives rise to a thermal bath of scalars and vectors in the dark sector with temperature $T_D/T\sim \sqrt{H_I/\Mpl}$, as computed in eq.~\eqref{eq:TD}. If the thermalization process is efficient 
the temperature of the dark sector can be as high as $H_I> f$ so that the symmetry remains unbroken until $T_D\sim f$.
 Assuming that at the phase transition the energy goes into the radial mode one finds
\begin{equation}\label{eq:thermal-coulomb}
\frac{\rho_\phi}s\bigg|_{\rm inf, thermal-PT} \sim f \left(\frac{H_I}{\Mpl}\right)^{3/2}  \longrightarrow \frac{\rho_A}s \sim \frac{M_A}{\sqrt{\lambda}} \left(\frac{H_I}{\Mpl}\right)^{3/2} 
\end{equation}
assuming again that the dark photon is lighter than the radial mode and non-relativistic today.

\begin{figure}
\begin{center}
\includegraphics[width=.65\textwidth]{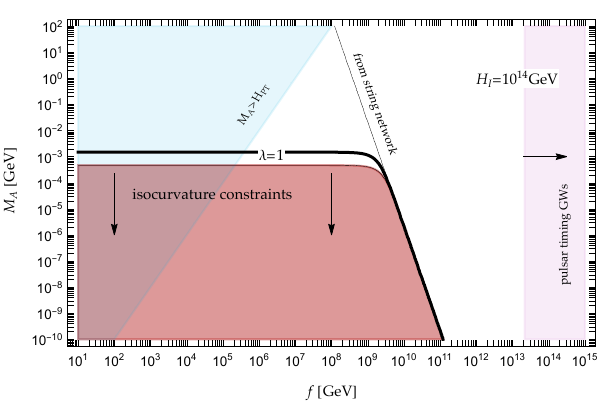}
\caption{\label{fig:coulomb}\small Coulomb phase of Dark Photon DM. DM abundance is reproduced along the solid black lines. Isocurvature constraints arise for small quartic couplings as discussed in the text. We also show the individual contribution to DM from the string network, with the dashed line where the dark photon mass is larger than Hubble at the phase transition. Above the line $M_A\gtrsim H_{\rm PT}$ contributions from the string network are suppressed.}
\end{center}
\end{figure}

In addition, the spontaneous U(1) also generates a string network. If the dark photon is lighter than Hubble at the phase transition 
the abundance of dark photons is again given by eq.~(\ref{eq:yield-strings}), which is independent on the details of the phase transition and does not depends explicitly on the dark sector temperature. 
The thermal contribution above can dominate over the string network if
$M_A >\lambda  f^4/H_I^3$.
Notice also that if the thermalization is instantaneous, the control parameter for the phase transition is the dark sector temperature rather than Hubble. This opens up the possibility that Hubble at the phase transition is smaller than the mass of the dark photon. Indeed $H_{\rm PT}\equiv H(T_D\approx f)\approx  f^2/H_I$, which by construction in this case is smaller than $f$. The emission of dark photons becomes inefficient for $H_{PT}< M_A$ that corresponds to
$g> f/H_I$. In this case the abundance is thus dominated by eq. (\ref{eq:abxi1H>f}) and \eqref{eq:thermal-coulomb}. 
A summary of the different contributions to the abundance is shown in figure \ref{fig:coulomb}.

\subsubsection{Isocurvature constraints}~\\
The abundance of dark photons (\ref{eq:abxi1H>f}) produced by inflationary fluctuations is IR dominated.
Since this is a significant or even dominant contribution to the total abundance it implies strong constraints from isocurvature perturbations\footnote{A related discussion arises for the QCD axion, 
see \cite{Lyth:1992tw,Rubakov}. For the QCD axion the isocurvature perturbations associated to the radial mode give a negligible contribution to the energy density of axion DM. This is not so in our case even if the symmetry is restored after inflation by thermal effects because the contribution (\ref{eq:abquartic}) grows with $f$.}.
As in section \ref{sec:isoHiggs} there are essentially two solutions to this problem. 
The first amounts to require that the isocurvature population is very subdominant compared to
the healthy one produced from cosmic strings or by the misalignment mechanism. The second solution is realized if the effective mass during inflation is comparable
to Hubble so that isocurvature perturbations on cosmological scales are suppressed. In this case the effective mass is controlled
by the quartic so this implies a lower bound of the quartic coupling \cite{Markkanen:2018gcw}. 

More in detail:
\begin{itemize}
\item $\lambda \ll 1$:\\
As discussed in the appendix \ref{sec:isocurvature} inflationary expansion makes the field $\Phi$ homogenous over a spatial distance,
\begin{equation}
d\sim \frac 1 {H_I} \exp[\sqrt{\frac {8\pi^2}{9\lambda}}]
\end{equation}
For $\lambda< 0.01$ this is larger than the size of the visible universe so that $\Phi$ has a constant value in our Hubble patch and the U(1) symmetry is broken. 
In particular if inflation lasts a number of e-foldings larger than $\sqrt{8\pi^2/(9\lambda)}$ the variance of the field approaches the constant value
\begin{equation}
\langle \phi^2 \rangle\approx \frac{H_I^2}{(2\pi)\sqrt{2\lambda}}
\label{eq:var}
\end{equation}
From this expression we can compute the isocurvature power as in eq.~\eqref{eq:iso-Delta}. Requiring that the isocurvature component is small requires $\lambda < 10^{-10}$, see also chapter 14 in \cite{Rubakov}.

\item $\lambda \sim 1$:\\
For large couplings $\Phi$ is dishomogeneous in the visible universe. This means that the symmetry is effectively restored during inflation 
and in particular a network of strings is formed when the symmetry breaks afterwards \cite{Lyth:1992tw}. Differently from the QCD axion this does not automatically solve the isocurvature problem because 
the contribution from inflationary fluctuations of $\Phi$ is significant. The modes of cosmological size are however suppressed if the effective mass $3\lambda \langle \phi \rangle^2$ is comparable with Hubble
where the variance is given by eq. (\ref{eq:var}). As discussed in the appendix B one finds that isocurvature constraints can be solved for $\lambda \sim 1$.
\end{itemize}

\section{Conclusions}

\begin{table}[h]
\begin{center}\small
\begin{tabular}{ c | c | c | c }
 \rowcolor{lightgray} & $M_\phi \gg H_I \gg M_A$ & $H_I \gg M_\phi \gg M_A$ & $H_I \gg M_A \gg M_\phi$ \\ \hline
\rowcolor{white} $f>H_I$  & GMR & GMR + $\phi$-decay + (iso) $\xi=0$ & GMR + thermal-FO + (iso) $\xi=0$\\ 
\rowcolor{white} $f<H_I, \xi = \frac16 $& -- & string network & string network \\ 
\rowcolor{white}$f<H_I, \xi = 0$& -- & string net.+ $\phi$-decay +(iso) & string net. + thermal-PT +(iso)   \\  
\end{tabular}
\caption{\label{tab:summary}\small  Summary of different phases of Dark Photon Dark Matter. The pure Dark Photon scenario
is only reproduced for $f>H_I$ and small gauge coupling. In other regions string network or thermal contribution to the abundance are often dominant. 
For minimal coupling strong isocurvature constraints generically arise.}
\end{center}
\end{table}

In this note we studied realizations of vector dark matter obtained through the Higgs mechanism.
Differently from the popular scenario where the mass is generated by the Stueckelberg mechanism this requires extra degrees of freedom 
and interactions. We focused in particular on the minimal construction with a dark abelian Higgs model
where a single scalar degree of freedom is added, the dark photon Higgs scalar. The scalar coupling to curvature is expected to capture more
general constructions such as dynamical symmetry breaking through strong interactions.

Already in this minimal realization a landscape of possibilities emerges, see table \ref{tab:summary}.
As for the QCD-axion two different scenarios exist depending on the scale of inflation.
Roughly, if the Hubble parameter during inflation is larger than the U(1) symmetry breaking 
scale $f$, the symmetry is restored during inflation so that the vector is effectively massless and not produced by
inflationary fluctuations. Dark photons are in this case generated through the decay of the scalars produced during inflation
or from cosmic strings.

In the opposite regime the symmetry is broken throughout the history of the universe so that
the vector is massive and produced during inflation similarly to Stueckelberg theories. However also the Higgs scalar
is typically produced during inflation and its decay changes the dark photon abundance. The scenario of Ref. \cite{Graham:2015rva}
is only reproduced for tiny gauge couplings that might be problematic from the point of view of the weak gravity conjecture.

An interesting aspect of the Stueckelberg dark photon is that it automatically avoids constraints from isocurvature perturbations produced during 
inflation also predicting effects for structures \cite{Gorghetto:2022sue,Amin:2022pzv}. In realizations through the Higgs mechanism we find that isocurvature constraints are typically re-introduced except if the coupling to curvature is conformal. In this case the approximate Weyl invariance of the action suppresses inflationary perturbations 
on cosmological scales.  

This motivates the study of inflationary production of dark sectors that are approximately Weyl invariant to automatically avoid isocurvature constraints. 
In this case the population of the dark sector might be dominated from the phase transition.
Another possibility is to change the dynamics during inflation so that minimally coupled scalars are not present 
at the beginning of inflation but emerge for example through a phase transition. We plan to return to these questions in future work.

{\small
\subsubsection*{Acknowledgements}
We would like to thank Lorenzo Ubaldi for collaboration at the initial stages of this project and Luca Di Luzio, Edward Hardy, Alessio Notari  and Giovanni Villadoro for discussions. 
This work is supported by MIUR grants PRIN 2017FMJFMW and 2017L5W2PT and INFN grant STRONG.
}

\appendix
\section{Power spectra of free fields}\label{app:A}
In this appendix we review the computation of power spectrum generated by inflation 
of minimally coupled scalar and massive vector fields, see \cite{Graham:2015rva,Ema:2018ucl,Ema:2019yrd,Ahmed:2020fhc,Kolb:2020fwh}.

To compute the energy density at late time we start from the power spectrum of fields at horizon exit, $k\approx a H_I$: in the linear regime
the energy density at later time will be the product of the primordial power spectrum times the `transfer functions' determined by the subsequent (classical) evolution. The spatial Fourier transform of a canonically normalized field $X(t,\vec{x})$ has a power spectrum
\be
\langle X(t,\vec{k}) X^*(t,\vec{q})\rangle = (2\pi)^3\delta^{(3)}(\vec k -\vec q)\,\, \frac{2\pi^2}{k^3} P_X(t,k)\,.
\ee
For the problem under consideration the power specrum is a combination of the power spectrum at horizon crossing and the later evolution, $P_X(t,k)\approx P_X(k)|_{\rm exit} \times |\mathrm{transfer}(t,\vec{k})|^2$, captured by a transfer function $T(t,\vec{k})$, $X(t,\vec k)\approx X(\vec k)|_{\rm exit}\times T(t,\vec{k})$ . These quantities are determined by the equation of motion with Bunch-Davies boundary conditions deep inside the horizon.

\subsection*{Scalar}
A scalar field of mass $M$ with a coupling to curvature $\xi$ has an equation of motion
\be\label{eq:scalarX}
\ddot X + 3 H \dot X + \frac{k^2}{a^2}X + (M^2+\xi R)X=0\,.
\ee
The above equation is used to determine the amplitude of the field at horizon exit, as well as the transfer function at later stages. The average energy density at any time can be expressed by exploiting the expression for the power spectrum of the Fourier transform $X(t,\vec k)$
\be\label{eq:rhoX}
\rho_X(t)= \langle \frac{\dot X^2}{2} +\frac{|\nabla X|^2}{2a^2} +M^2 \frac{X^2}{2}\rangle \approx \int d \log k\, \frac{P_X(k)|_{\rm exit}}{2}\, \left[|\dot T(t,\vec k)|^2+(\frac{k^2}{a^2}+M^2)|T(t,\vec k)|^2\right]\,.
\ee
The above expression is then easily computed by knowing the primordial power spectrum (using Bunch-Davies boundary conditions) and the transfer function $T(t,\vec k)$. The power spectrum at horizon exit (and hence on super-horizon scales) is given by
\be
P_X(k)|_{\rm exit}=\frac{H_I^2}{(2\pi)^2} %\left( \frac{k}{a H_I}\right)^{3-2\sqrt{\frac{9}{4}-12\xi-\frac{M^2}{H_I^2}}}\,.
\ee
The transfer function $T(t,\vec k)$ instead can be derived in simple scaling limits, such as
\be\label{eq:transferX}
T_{\xi=0}(t,\vec k)\approx \left\{\begin{array}{lcl} 1 & & k/a\ll  H\,\, \mathrm{and}\,\, M \ll H\,,\\
\exp(i M t)/a^{3/2} & & k/a \ll M\,\, \mathrm{and}\,\, H \ll M\,,\\
\exp(i k/a\, t)/a & & M, H \ll k/a\,.
\end{array}
\right.
\ee

From this expression we can compute the energy density today per wave-number that we define $d \rho_X/d\log k$, which corresponds to the function under the integral sign in \eqref{eq:rhoX}. Each mode will behave according to \eqref{eq:transferX} when it enters the relevant regimes. It is convenient to define $k_{\rm eq}=a_{\rm eq} H(a_{\rm eq})$, $k_*\equiv a(H=M)M\equiv a_* M$, and also $k_{\rm max}=a_{\rm end} H_I$. Since we can roughly estimate $k_{\rm max}\approx T_{0} \sqrt{\Mpl/T_R}$, for the model under consideration we have $M\gg k_{\rm max}$, such that today all the modes are non-relativistic. Moreover, we define $a_k$ the scale factor when the mode $k$ re-enters the horizon: $a_k H(a_k)=k$.
\be
\frac{d \rho_X}{d \log k}\bigg|_{\rm today}\approx P_X(k)|_{\rm exit}\left\{\begin{array}{lcl} 
\displaystyle \frac{k^2}{a_k^2} \frac{a_k^4}{a(k=aM)^4}\frac{ a(k=aM)^3}{a_0^3}= M k a_k^2 \approx H_{\rm eq}^2 \frac{M}{k} a_{\rm eq}^4  & & k\gg k_*\,,\\
\displaystyle M^2 \frac{a_*^3}{a_0^3}\approx H_{\rm eq}^2 \sqrt{M/ H_{\rm eq}} a_{\rm eq}^3  & & k \ll k_*\,,\\
\end{array}
\right.
\ee
where $k_*\approx a_{\rm eq}\sqrt{H_{\rm eq} M}$. Since $k_*\gg k_{\rm eq}$ we do not expect a qualitative different behavior for modes that re-enters during matter domination. Normalizing to the peak infra-red part of the spectrum we get, and making use of the relation
\begin{equation}
H_*^2 =\frac{ \pi^2}{90} g_* \frac{T_*^4}{\Mpl^2} \longrightarrow s_* = \frac {2\pi^2}{45}g_* T_*^3= 2.3  g_*^{1/4} (H_* \Mpl)^{3/2},
\label{eq:sH}
\end{equation}
we get
\be
\frac 1 s \frac{d \rho_X}{d \log k}\big|_{\xi=0}\approx \frac {1}{g_*(M)^{1/4}} \frac{H_I^2}{(2\pi)^2} \frac{\sqrt{M}}{\Mpl^{3/2}}  \left\{\begin{array}{lcl} 
\displaystyle   \frac{k_*}{k} & & k\gg k_*\,,\\
\displaystyle 1  & & k \ll k_*\,,\\
\end{array}
\right.\,,
\ee

\subsection*{Vector}

The computation for the vector field is very similar. A Stueckelberg vector $A_\mu(t,\vec{k})$ with mass $M$ has the following equation of motion for the longitudinal polarization \cite{Graham:2015rva}
\be
\ddot A_L +  H \left(1+\frac{2k^2}{k^2 + a^2 M^2}\right) \dot A_L + \frac{k^2}{a^2} A_L + M^2 A_L=0\,,
\ee
which in the relativistic limit ($k\gg a M$) resembles the equation for a minimally coupled scalar \eqref{eq:scalarX}.
The corresponding energy density is given by
\be
\rho_A(t) = \frac{1}{a^2} \int d\log k \frac{P_X(k)|_{\rm exit}}{2} \frac{k^2}{M^2}\left[\frac{a^2 M^2}{a^2 M^2 + k^2} |\dot T|^2 + M^2 |T|^2 \right]
\ee
where we have normalized the power spectrum to the case of a minimally coupled scalar $X$ described before (in this case it is the Goldstone eaten by the vector). In the above expression $T$ is again the transfer function for the field $A_L$ with trivial initial conditions. In the relevant scaling limits, the expression for the transfer function is again given by \eqref{eq:transferX} with a slight modification\footnote{Continuity of the first derivative across the matching condition, ensure that the growing solution for $T$ in the region $k/a\ll  H\,, M \ll H$, where the vector equation of motion differs from the scalar case, is absent  \cite{Graham:2015rva}.} 
\be\label{eq:transferL}
T_{L}(t,\vec k)\approx \left\{\begin{array}{lcl} 1 & & k/a\ll  H\,\, \mathrm{and}\,\, M \ll H\,,\\
\exp(i M t)/\sqrt{a} & & k/a \ll M\,\, \mathrm{and}\,\, H \ll M\,,\\
\exp(i k/a\, t)/a & & M, H \ll k/a\,.
\end{array}
\right.
\ee
As compared to the case of the scalar we see that when the mass term dominates on super-horizon scales, where $T_L\approx 1$, the energy density redshifts as $1/a^2$. This behaviour is instrumental to suppress long modes contributions to the power spectrum today (see figure \ref{fig:PW} for an explicit calculation). By direct inspection we arrive at the following decomposition of the power spectrum
\be
\frac 1 s \frac{d \rho_A}{d \log k}\big|_{L}\approx \frac {1}{g_*(M)^{1/4}} \frac{H_I^2}{(2\pi)^2} \frac{\sqrt{M}}{\Mpl^{3/2}}  \left\{\begin{array}{lcl} 
\displaystyle  k_*/k & & k\gg k_*\,\\
\displaystyle 1  & & k \approx k_*\,\\
\displaystyle k^2/k_*^2  & & k \ll k_*\,\\
\end{array}
\right.\,.
\ee

\section{Isocurvature perturbations}
\label{sec:isocurvature}
In this appendix we derive the contribution to isocurvature perturbations for a minimally coupled scalar with non vanishing mass and quartic coupling.

When the mass or the quartic coupling are sizable, inflationary perturbations of a minimally coupled scalar are suppressed for the modes that exit at the beginning of inflation
corresponding to cosmological scale in our universe today \cite{Markkanen:2018gcw,Markkanen:2019kpv,Tenkanen:2019aij}.
This can be seen as follows. The power spectrum of a massive scalar field during inflation reads
\begin{equation}
P(k)= \frac{2\pi^2}{k^3}\Delta(k)= \frac {H_I^2} {2k^3} \left( \frac k {a H_I}\right)^{3-2\nu}\,,~~~~~~~~\nu= \sqrt{\frac 9 4- \frac {M^2}{H_I^2}}\approx \frac 3 2 -\frac {M^2}{3H_I^2}\,.
\end{equation}
The power spectrum allows us to compute the equal-time 2-point function as $\langle \phi(\vec x)\phi(\vec y)\rangle = \int d^3k/(2\pi)^3 P(k)e^{-i\vec{k}\cdot(\vec x- \vec y)}$. Therefore the variance of the field at a point in space is given by the following quantity (at the end of inflation)
\begin{equation}
\langle \phi^2(\vec x) \rangle = \frac {H_I^2}{(2\pi)^2} \int_{H_I a_e e^{-N_I}}^{H_I a_e} \frac {dk}k \left( \frac k {a_e H_I}\right)^{\frac {2M^2}{3H^2}}\approx 
\left\{\begin{array}{lcl} 
\displaystyle  \frac{3 H_I^4}{8\pi^2 M^2}\,. & & N_I > \frac{H_I^2}{M^2}\,,\\
\displaystyle N_I \frac {H_I^2}{(2\pi)^2}  & & N_I < \frac{H_I^2}{M^2}\,,\\
\end{array}
\right.\,,
\label{eq:vev}
\end{equation}
where $a_e$ is the scale factor at the end of inflation and $N_I$ is the total number of e-foldings of inflation. 
The fact that $\langle \phi^2 \rangle$ approaches a constant value is due to the balance between the stochastic inflationary fluctuations and the
the classical force relaxing the field to the minimum. From the point of view of the power spectrum the convergence of the integral in the IR is due to the suppression of long modes
 that thus cease to contribute to $\langle \phi^2 \rangle$. 

The observation above has an important consequence when   $H_I^2/M^2$ is smaller than the number of e-foldings of visible inflation because 
modes of cosmological size today are suppressed. This allows to avoid isocurvature constraints \cite{Tenkanen:2019aij}.

We can explicitly compute the power spectrum of the isocurvature DM perturbations $P_{\rm iso}(k)=(2\pi^2/k^3)\Delta_{\rm iso}(k)$ from the definition of the 2-point function of the density contrast $\delta\rho/ \bar \rho$,
\be
\frac {\langle\delta \rho(x)\delta \rho(0)\rangle}{\langle \rho \rangle^2}= \int \frac{d^3k}{(2\pi)^3} P_{\rm iso}(k) e^{-i \vec k \cdot \vec x} \,.
\ee 
Given that the energy density is proportional to  $\rho(x)\propto\phi^2(x)$, and under the assumption of gaussianity, we have
\begin{equation}
 \frac {\langle\delta \rho(x)\delta \rho(0)\rangle}{\langle \rho \rangle^2}  = \frac {\langle\phi^2(x)\phi^2(0)\rangle -\langle \phi^2\rangle^2}{\langle \phi^2\rangle^2} = 2\frac{\langle \phi(x)\phi(0)\rangle^2}{\langle \phi^2 \rangle^2}
\end{equation}
The isocurvature spectrum is therefore related to the square of the power spectrum of $\phi$. By taking the inverse Fourier transform, we get
\be\label{eq:iso-Delta}
\Delta_{\rm iso}(k)=\frac {k^3}{2\pi^2}\frac 2 {\langle \phi^2\rangle^2} \int d^3x e^{i \vec k \cdot \vec x} \langle\phi(x)\phi(0)\rangle^2 =\frac{k^3}{2\pi^2}\frac{2}{\langle \phi^2\rangle^2} \int \frac{d^3q}{(2\pi)^3}P(q)P(k-q)\,.
\ee
The above integral can be computed directly\footnote{
Given that $P$ is a power low in the wave-number of the form $P(Q)\propto(Q^2)^{-\alpha}$, we have
\begin{eqnarray}
\int \frac{d^3q}{(2\pi)^3}P(q)P(k-q)&=&A^2 \frac{\Gamma(2\alpha)}{\Gamma(\alpha)^2}\int_0^1 dx \int \frac{d^3Q}{(2\pi)^3}\frac{x^{\alpha-1}(1-x)^{\alpha-1}}{[Q^2+ k^2 x(1-x)]^{2\alpha}},\\
&=&A^2 \frac{\Gamma(2\alpha)}{\Gamma(\alpha)^2}\,\frac{k^{2(n_\phi-1)}}{k^3} \frac{1}{(4\pi)^{3/2}}\frac{\Gamma(2\alpha-3/2)}{\Gamma(2\alpha)}\int_0^1 dx \frac{x^{\alpha-1}(1-x)^{\alpha-1}}{[ x(1-x)]^{2\alpha-3/2}}\\
&=&A^2 \frac{\Gamma(2\alpha)}{\Gamma(\alpha)^2}\,\frac{k^{2(n_\phi-1)}}{k^3} \frac{1}{(4\pi)^{3/2}}\frac{\Gamma(2\alpha-3/2)}{\Gamma(2\alpha)}\, 2^{2\alpha-2}\sqrt{\pi}\frac{\Gamma(\frac32-\alpha)}{\Gamma(2-\alpha)}\end{eqnarray}
with $2\alpha=3-(n_\phi -1)$ and $A^2=H_I^4/4 (1/a_e H_I)^{2(n_\phi-1)}$, where $n_\phi$ the tilt of the power spectrum.}. For $N_I> H_I^2/M^2$ using eq. (\ref{eq:vev})  one finds,
\be\label{eq:iso}
\Delta_{\rm iso}(k)\approx \frac {8M^2}{3H_I^2} \left( \frac k {a_e H_I}\right)^{\frac {4M^2}{3H_I^2}}
\ee
From this expression we can see explicitly that cosmological modes, that exit at the beginning of inflation, have a suppressed power.  
In particular, the mode $k_*$ that exits the horizon during the last $N_*$ e-fold of inflation experiences a suppression
\begin{equation}
\Delta_{\rm iso}(k_*) \approx \frac {8M^2}{3H_I^2} e^{- \frac{4 N_I M^2}{3H_I^2}}\,.
\label{eq:iso-star}
\end{equation}
For $N_I\approx 50$, a value consistent with the fact that the field has already reached the stationary state \eqref{eq:vev}, we can estimate the strongest bound arising from isocurvature perturbations. 
At cosmological scales tested by the CMB it amounts to have $\Delta_{\rm iso}/\Delta_\zeta|_{k_*} \lesssim 0.035$. Satisfying this bound implies
\begin{equation}
 M \gtrsim 0.5 H_I\, \quad \quad \mathrm{(isocurvature)}
\end{equation}
Such value is still compatible with the assumption that the field is light during inflation and can be efficiently produced via inflationary fluctuations.

A different situation arises when $H_I\gg f$ because the quartic coupling dominates the potential. For this situation, the previous formulas cannot be directly applied to assess the impact of isocurvatures 
and one should properly use the formalism of stochastic inflation  \cite{Starobinsky:1994bd}. An approximate result can be obtained by expanding the potential around the classical value of $\phi$. Doing so, the second derivative of the scalar potential appears in the equation of motion for the fluctuations. Since $V''= \lambda(3 \phi^2- f^2)$ for $\phi\sim H_I$ the effective mass is $M^2\approx 3 \lambda \langle \phi^2\rangle$. By plugging this value into \eqref{eq:vev}  we get
\begin{equation}
\langle \phi^2 \rangle\approx \frac{H_I^2}{(2\pi)\sqrt{2\lambda}}
\end{equation}
a value that is reached after a number of e-foldings $N_I\sim \sqrt{8\pi^2/9\lambda}$.
As before the variance does not grow in time so that long wave-length modes are suppressed. Indeed, even for vanishing masses, the evolution of the field variance on super-horizon scales reaches a steady configuration set by the compensation between random walk and the quartic interaction \cite{Starobinsky:1994bd}. Following the steps that brought us to \eqref{eq:iso} we get 
\begin{equation}
\Delta_{\rm iso}|_{\rm quartic} (k_*)\approx\frac{2\sqrt{2\lambda}}{\pi} e^{- \frac{N_* \sqrt{2\lambda}}{\pi}}\,.
\label{eq:iso-quartic}
\end{equation}
Applying the same bound from the CMB, we get a lower limit on the size of the quartic coupling in order to satisfy the constraints from isocurvatures
\begin{equation}
 \lambda \gtrsim 1\, \quad \quad \mathrm{(isocurvature)}\,.
\end{equation}
where we assumed 50 e-folding of visible infllation.
A more precise estimate can be obtained using the stochastic formalism of inflation \cite{Starobinsky:1994bd}.
In that case one finds $\Delta_{\rm iso}\approx 1.5 \sqrt{\lambda} e^{-0.58 N_* \sqrt{\lambda}}$ giving similar constraints on $\lambda$.

\pagestyle{plain}
\bibliographystyle{jhep}
\small
\bibliography{biblio}

\end{document}